\documentstyle{article}

\begin{document}

\LaTeX{}\bigskip\ \bigskip\ \bigskip\

\begin{center}
Predicting phase transition pressure in solids:a semi-classical possibility%
\medskip\

V.Celebonovic\smallskip\

Institute of Physics,Pregrevica 118,11080 Zemun-Beograd,Yugoslavia

vladan@phy.bg.ac.yu

vcelebonovic@sezampro.yu\bigskip\
\end{center}

Abstract: This is a short review of the physical ideas,algorithm for
calculations of the phase transition pressure and some results of a
semi-classical theory of the behaviour of materials under high pressure
proposed by P.Savic and R.Kasanin.It is based on the Coulomb interaction
supplemented by a microscopic selection rule and a set of experimentally
founded postulates.The theory has been applied to cases ranging from DAC
experiments to calculations of models of planetary internal structure.%
\newpage\

\begin{center}
Introduction\smallskip\
\end{center}

The study of materials under high pressure (and temperature) is important in
a variety of situations in physics,astrophysics and related sciences.These
range from highly exotic examples such as the early Universe, to laboratory
experiments performed in diamond anvil cells (DACs). Determining
theoretically phase diagrams of solids under high pressure is an extremely
complex problem in statistical mechanics (for example [1] ).It involves the
choice of a proper Hamiltonian of the system under consideration,the
calculation of the free energy,and then the determination of the regions (or
points) in the parameter space in which the thermo-dynamical potentials
become non-analytical functions.

The purpose of this paper is to present briefly the main physical ideas and
examples of applicability of a particular semi-classical theory of the
behaviour of materials under high pressure.It was proposed by.P.Savic and
R.Kasanin [2] and nicknamed the SK theory for short.

Initially,the idea for work on this problem dates back to a paper by Savic
[3],which had the aim of exploring the origin of rotation of celestial
bodies.It emerged from this work that rotation is closely related to the
internal structure,and that a theory of the behaviour of materials under
high pressure was needed to explain it.The theory which he developed with
Kasanin proposes a simple algorithm for the calculation of phase transition
pressure in materials subdued to pressure.Recent work [4] gives the
possibility of establishing the equation of state and calculating the bulk
modulus and its first pressure derivative.Apart from laboratory studies,the
SK theory has found applications in astrophysics ( such as [5]-[9] and
references given there).Note that astrophysical results close to those
obtained within SK have been derived by Vasiliev [10] using different
methods.

This paper has two more sections: the next one contains an outline of the
basic premises of the SK theory,while the third one is devoted to a brief
presentation of applications of this theory to laboratory experiments.%
\newpage\

\begin{center}
Basic premises of the SK theory\medskip\
\end{center}

The object of study in the SK theory is a mole of any material under high
external pressure.SK is based on the idea that increasing pressure leads to
excitation and ionisation of atoms and molecules that make up the
material.Translated into quantum mechanical terms,this means that increased
pressure provokes the expansion of the radial part of the electronic wave
functions in the atoms and molecules that make up the material.Such an idea
might seem strange at first sight,because one may be inclined to think that
high external pressure leads to a ''crunch''.However,this problem has
received a quantum mechanical treatment only about a decade ago [11] in a
series of papers aiming at a theoretical explanation of the ruby scale. The
mean inter-particle distance $a$ is defined in the SK theory by the relation

\begin{equation}
\label{(1)}N_A(2a)^3\rho =A
\end{equation}

where $N_A$ is Avogadro's number,$\rho $ the mass density and $A$ the mean
atomic mass of the material.One can now define the ''accumulated'' energy
per electron as

\begin{equation}
\label{(2)}E=e^2/a
\end{equation}

It can be shown that $a$ as defined above is a multiple of the Wigner-Seitz
radius [12].

Mathematically,the basic premises of the SK theory are expressed in a series
of 6 postulates which establish the ratio of densities and accumulated
energies in successive phases [6]-[8]. Due to the fact that these postulates
assume that jumps of the density occur in phase transitions,the
applicability of the SK theory is limited to first order phase
transitions.In all calculation within SK an integer index (denoted by $i$ )
appears.Physically,it has the meaning of the ordinal number of a first order
phase transition occuring in a material.The phase indexed by $i=0$ ends in
the critical point.Without entering into details,the main practical result
is the following expression for the phase transition pressure $p_{tr}$

\begin{equation}
\label{(3)}p_{tr}=\{_{0.6785p_i^{*};i=2,4,6,..}^{0.5101p_i^{*};i=1,3,5,...}
\end{equation}
\newpage\ where

\begin{equation}
\label{(4)}p_i^{*}=1.8077\beta _i\left( V\right) ^{-4/3}2^{4i/3}Mbar
\end{equation}
\smallskip\
\begin{equation}
\label{(5)}\beta _i=3\left( \alpha _i^{1/3}-1\right) /\left( 1-1/\alpha
_i\right)
\end{equation}

and
\begin{equation}
\label{(6)}\alpha _i=\{_{5/3;i=2,4,6,..}^{6/5;i=1,3,5,..}
\end{equation}
The symbol V denotes the molar volume of the material under standard
conditions,and $i$ is an integer index numbering first order phase
transitions which occur in a material.Physically realizable values of this
index are determined by the selection rule

\begin{equation}
\label{(7)}E_0^{*}+E_I=E_i^{*}
\end{equation}
\smallskip\ where $E_I$ is the ionisation or excitation potential and $%
E_0^{*}$ and $E_i^{*}$ are pressure dependent characteristic energies of the
specimen,which take into account only the Coulomb part of the inter-particle
interaction potential.In another line of development,recent work [4] has
given expressions for the equation of state and implicitly for the bulk
modulus and its first pressure derivative within the SK theory.It was shown
there that the equation of state has the form:

\begin{equation}
\label{(8)}P\cong \frac{2e^2}3\left( \frac{N_A}A\right) ^{1/3}\rho
^{4/3}\left[ C\frac{N_A}A+B\rho \exp \left[ 4W\left( 1-\left( \frac \rho
{\rho ^{*}}\right) ^{1/3}\right) \right] +...\right]
\end{equation}

\smallskip\ where $N_A$ denotes Avogadro's number,$A$ is the mean atomic
mass of the material,$\rho $ the mass density and other symbols denote
various constants calculable within SK.\newpage\

\begin{center}
Applications in laboratory experiments
\end{center}

The algorithm proposed by SK for the calculation of the phase transition
pressure was applied to 22 different materials [7],[8].Apart from 20
materials for which values of phase transition pressure were known
experimentally or from various theoretical calculations $%
(C_6H_6;CH_4;CCl_4;C_{60};CsI;CF_2\equiv
CF_2;CdS;Al;Ba;CaO;Kr_2;Ne;LiH;Te;TeO_2;Mg_2SiO_4;Pb;S_2;AlPO_4;$

$GdAlO_3:Cr^{3+})$ the SK algorithm was applied to hydrogen and helium.
They were included in the calculation because of their astrophysical
importance,although the actual existence of phase transitions under
high pressure in these two materials has not yet been confirmed in
static high pressure experiments.
However,metallic-like behaviour was obtained in shock-compression
work at a pressure lower than the theoretically predicted value
[13],[14]. The relative discrepancies between the values of the phase
transition pressure calculated within SK and those existing in the
literature vary between nearly 0\% and 30\%. These differences are
due to a variety of factors.

For example,the precision of experimental data and of the input parameters
in the calculations account for approximately $\pm 10\%.$An important source
of the discrepancies is the form of the interparticle potential.The SK takes
into account only the pure Coulomb part of the interparticle
potential,without the contributions of the charge distribution overlap and
of the dispersive and repulsive forces.Now,the relative contribution of the
dispersive and repulsive forces to the full inter-particle potential is
minimal for C,H,N,O.Interestingly,the discrepancies between the SK and
experimental values of the phase transition pressure is also minimal for the
hydrocarbons [7].

Taking into account the existence of the charge distribution overlap gives
rise to three additional terms in the expression for the accumulated
energy.The sum of these terms is pressure dependent and it can be
positive,negative or zero [7] .The existence of these terms induces an error
in the calculated values of the phase transition pressure,and the magnitude
of this error is also pressure dependent.Due to space limitations,we have
here outlined the influence of just two factors which contribute to the the
relative discrepancies between the SK values of the phase transition
pressure in various materials and those obtained experimentally.A detailed
analysis is avaliable in [7].

Instead of a conclusion,several comments about the SK theory see to be in
order.The big advantage of this theory is its physical and calculational
simplicity.On the other hand,this simplicity necessarily induces
discrepancies between the experimental values of the phase transition
pressure and those calculated within SK.Pushing the reasoning further,the
existence of these discrepancies opens up the possibilities for improving
the basic assumptions of SK by rendering them more complex and physically
realistic.\newpage\ One of the possible directions for improving the theory
would be the inclusion of non-Coulombic terms into the expression for the
accumulated energy.Some work along these lines has already started.

\begin{center}
Acknowledgement
\end{center}

The author is grateful to MSTS for financial support.

\begin{center}
References
\end{center}

[1] J.W.Negele and H.Orland:Quantum Many-Particle Systems [Addison\\ Wesley Publ.Comp.,New York,1988].

[2] P.Savic and R.Kasanin:\ The Behaviour of Materials Under High Pressure
I-IV [SANU,Beograd,1962/65].

[3] P.Savic,Sur l'origine de la rotation des corps celestes,Bull.de la
classe des Sc.Math. et Natur. de l'Acad.Serbe des Sci.et des Arts,{\bf XXVI}%
,107 (1962) .

[4] V.Celebonovic, preprint astro-ph/9910457, prepared for the XII National
Conference of Yugoslav Astronomers,Belgrade,November 1999.

[5] P.Savic,The internal structure of the planets Mercury,Venus,Mars and
Jupiter according to the Savic-Kasanin theory,Adv.Space Res.,{\bf 1},131
(1981) .

[6] V.Celebonovic,Hydrogen and helium under high pressure:a case for a
classical theory of dense matter,Earth,Moon and Planets,{\bf 45},291 (1989d).

[7] V.Celebonovic,High pressure phase transitions-examples of classical
predictability,ibid,{\bf 58},203 (1992c).

[8] V.Celebonovic,The origin of rotation,dense matter physics and all that:
a tribute to Pavle Savic,Bull.Astron.Belgrade,{\bf 151},37 (1995).

[9] P.Savic and V.Celebonovic,Dense matter theory:a simple classical
approach,AIP Conf.Proc.,{\bf 309} , 53 (1994).

[10] B.V.Vasiliev,Thomas Fermi Calculation of Gravity Induced Electrical
Polarization in Dense Electron-Nuclear Plasma,Dubna preprint P17-99-76\\
(1999).

[11] D.Ma,Z.Wang,J.Chen and Z.Zheng,Theoretical calculations of pressure
induced blue and red shifts of spectra of ruby,J.Phys.,{\bf C21},3585 (1988).

[12] Y.C.Leung,Physics of Dense Matter,[Science Press/World Scientific,\\Beijing and Singapore,1984].

[13] S.M.Pollaine and W.J.Nellis,Laser-Generated Metallic Hydrogen,LLNL
preprint UCRL-JC-135176 (1999).

[14] W.J.Nellis,Hydrogen at High Pressure and Temperature,LLNL preprint
UCRL-JC-135910 (1999).

\end{document}